\documentclass[aps,prl,superscriptaddress,reprint]{revtex4-1}

\usepackage{graphicx}
\begin{document}


\newcommand{\CBST}{Cr$_{0.08}$(Bi$_x$Sb$_{1-x}$)$_{1.92}$Te$_3$}
\newcommand{\CBSTA}{Cr$_{0.08}$(Bi$_{0.37}$Sb$_{0.63}$)$_{1.92}$Te$_3$}
\newcommand{\BST}{(Bi,Sb)$_2$Te$_3$}
\newcommand{\BS}{Bi$_2$Se$_3$}
\newcommand{\BT}{Bi$_2$Te$_3$}
\newcommand{\TC}{$T_{\rm{C}}$}
\newcommand{\Vsp}{$V_{\rm{SP}}$}
\newcommand{\jc}{$\vec{j}_c$}
\newcommand{\Jc}{$\vec{J}_c$}
\newcommand{\Js}{$\vec{J}_s$}
\newcommand{\Rxx}{$R_{xx}$}
\newcommand{\DHpp}{$\Delta H_{\rm{PP}}$}
\renewcommand{\vec}[1]{\mathbf{#1}}

\title{Fermi Level Dependent Spin Pumping from a Magnetic Insulator into a Topological Insulator}


\author{Hailong Wang}
\affiliation{Department of Physics and Materials Research Institute, the Pennsylvania State University, University Park, PA 16802, USA}
\author{James Kally}
\affiliation{Department of Physics and Materials Research Institute, the Pennsylvania State University, University Park, PA 16802, USA}
\author{C\"{u}neyt \c{S}ahin}
\affiliation{Department of Physics and Astronomy, Optical Science and Technology Center, University of Iowa, Iowa City, Iowa 52242, USA\\Institute for Molecular Engineering, University of Chicago, Chicago, IL 60637, USA}
\author{Tao Liu}
\affiliation{Department of Physics, Colorado State University, Fort Collins CO 80523}
\author{Wilson Yanez}
\affiliation{Department of Physics and Materials Research Institute, the Pennsylvania State University, University Park, PA 16802, USA}
\author{Eric J. Kamp}
\affiliation{Department of Physics and Materials Research Institute, the Pennsylvania State University, University Park, PA 16802, USA}
\author{Anthony Richardella}
\affiliation{Department of Physics and Materials Research Institute, the Pennsylvania State University, University Park, PA 16802, USA}\author{Mingzhong Wu}
\affiliation{Department of Physics, Colorado State University, Fort Collins CO 80523}
\author{Michael E. Flatt\'e}
\affiliation{Department of Physics and Astronomy, Optical Science and Technology Center, University of Iowa, Iowa City, Iowa 52242, USA\\Institute for Molecular Engineering, University of Chicago, Chicago, IL 60637, USA}
\author{Nitin Samarth}
\email[]{nsamarth@psu.edu}
\affiliation{Department of Physics and Materials Research Institute, the Pennsylvania State University, University Park, PA 16802, USA}



\date{\today}

\begin{abstract}
Topological spintronics aims to exploit the spin-momentum locking in the helical surface states of topological insulators for spin-orbit torque devices. We address a fundamental question that still remains unresolved in this context: does the topological surface state alone produce the largest values of spin-charge conversion efficiency or can the strongly spin-orbit coupled bulk states also contribute significantly? By studying the Fermi level dependence of spin pumping in topological insulator/ferrimagnetic insulator bilayers, we show that the spin Hall conductivity is constant when the Fermi level is tuned across the bulk band gap, consistent with a full bulk band calculation. The results suggest a new perspective, wherein ``bulk-surface correspondence'' allows spin-charge conversion to be simultaneously viewed either as coming from the full bulk band, or from spin-momentum locking of the surface state. 
\end{abstract}


\maketitle

The helical Dirac surface states found in topological insulators (TIs)\cite{hasan2010,Heremans:2017aa} have attracted significant attention recently for potential applications in spintronics. This has led to a burgeoning field, topological spintronics, founded on the central concept that these topological surface states might provide a natural way to efficiently convert charge currents to spin currents \cite{Fan:2014aa,Mellnik:2014aa,Jamali:2015aa,Liu:2015a6,Deorani:2014aa,Wang:2015aa,Lee:2015aa,Rojas-Sanchez:2016aa,Kondou:2016aa,Wang:2016ay,Jiang:2016aa,Lv:2018aa}. Indeed, a series of experiments studying TI/FM bilayers have demonstrated tantalizingly large values of the spin torque ratio, the relevant figure of merit, even at room temperature \cite{Mellnik:2014aa}, thus setting the stage for spin-orbit torque devices that rely on all-electrical switching of a ferromagnet (FM)  \cite{Han:2017aa,Wang:2017bb,Mahendra:2018,Khang:2018}. An important question that remains to be resolved in this context is to separate the contributions of the surface states from those of the bulk states since they both have strong spin-orbit coupling. We note that many experiments measuring spin-charge conversion in TI/FM bilayers use TI layers wherein the chemical potential (or Fermi level) $E_F$ is in the bulk bands. Two recent experiments have studied spin-charge conversion as a function of $E_F$ in TI/FM heterostructures and find an anomaly when $E_F$ crosses the Dirac point \cite{Kondou:2016aa,Jiang:2016aa}. 

Here, we use spin pumping \cite{Tserkovnyak:2005aa,Heinreich2011,Mosendz,Wang:2014aa} in epitaxial \CBST~ thin films grown on the ferrimagnetic insulator Y$_3$Fe$_5$O$_{12}$ (YIG) to thoroughly study this problem. We vary $E_F$ in these \CBST~ films using both electrical gating in a single sample and compositional changes in a series of samples. Our experiments show that the spin Hall conductivity is constant when $E_F$ lies within the bulk band gap. In contrast to the measurements reported recently \cite{Kondou:2016aa,Jiang:2016aa}, our experiments indicate that there is no anomaly in the spin-charge conversion efficiency when $E_F$ crosses the Dirac point. We note that these prior experiments have key differences from the study presented here. The first experiment \cite{Kondou:2016aa} studied charge-to-spin conversion rather than spin-to-charge conversion. Further, the experiment used a more complex interface by inserting a thin metallic Cu layer between the topological insulator and a metallic ferromagnet. The second experiment \cite{Jiang:2016aa} uses a very different spin-to-charge conversion mechanism (spin Seebeck effect). In contrast, we use a clean interface with an insulating ferrimagnet that should preserve the chiral states\cite{Zhang2016} and a measurement approach (spin pumping) that is readily compared to theory since the measured signal (\Vsp) is directly connected to the spin Hall conductivity $\sigma_S \propto 1/V_{\rm{SP}}$. To explain our data, we calculate the bulk spin Hall conductivity for \BST~ using a Kubo formalism previously applied to another topological insulator (Bi$_x$Sb$_{1-x}$)~ \cite{Sahin:2015aa}. In this approach, we view the spin-charge conversion in a topological insulator from a ``bulk-surface correspondence'' perspective, similar to that in the quantum Hall effect \cite{Weis:2011aa}, arguing that the roles of the bulk states and helical surface states in spin-charge conversion cannot be separated, but should rather be viewed as equivalent. When the chemical potential is tuned into the conduction or valence band, the dramatic variation of the spin-charge conversion heavily depends on the charge carrier type, confirming the strong spin-orbit coupling of bulk states, which is of the same order as the inverse Rashba-Edelstein effect (IREE) measured when the chemical potential is in the bulk band gap. In contrast to simple models of topological insulators, in which the signs of the spin-orbit correlation of conduction and valence states near the band edge are opposite, here the signs of the spin-orbit correlation of bulk states of both the conduction and valence bands near the band edge are the same, leading to a plateau (rather than a minimum or maximum) spin Hall conductivity in the gap. Our experiments qualitatively (and, to some extent, quantitatively) confirm this theoretical perspective.

We grew 20-nm-thick YIG films on single-crystal Gd$_3$Ga$_5$O$_{12}$ (111) substrates by sputtering \cite{Chang:2014aa} followed by deposition of \CBST~layers using molecular beam epitaxy. The purpose of the slight Cr doping is to improve the crystalline quality of the thin film; the details of the growth method have been reported previously \cite{Wang:2016ay}. We first discuss the structural and interfacial characterization of the YIG/TI heterostructures. A representative $\theta - 2\theta$ x-ray diffraction (XRD) scan of a 40-quintuple- layer (QL)\CBSTA~film (Fig. 1(a)) indicates a phase-pure TI layer. A zoom-in view of the (003) peak curve in the inset to Fig. 1(a) exhibits pronounced Laue oscillations, indicating a smooth surface and a relatively sharp YIG/TI interface. The atomic ordering of the \CBST~ surface is also confirmed by the reflection high-energy electron diffraction pattern (supplementary materials). The atomic force microscopy (AFM) image of a YIG/\CBSTA~ (10 QL) bilayer in Fig. 1(b) gives a root-mean-square (rms) roughness of about 1 nm.

\begin{figure}
\includegraphics[width=3in]{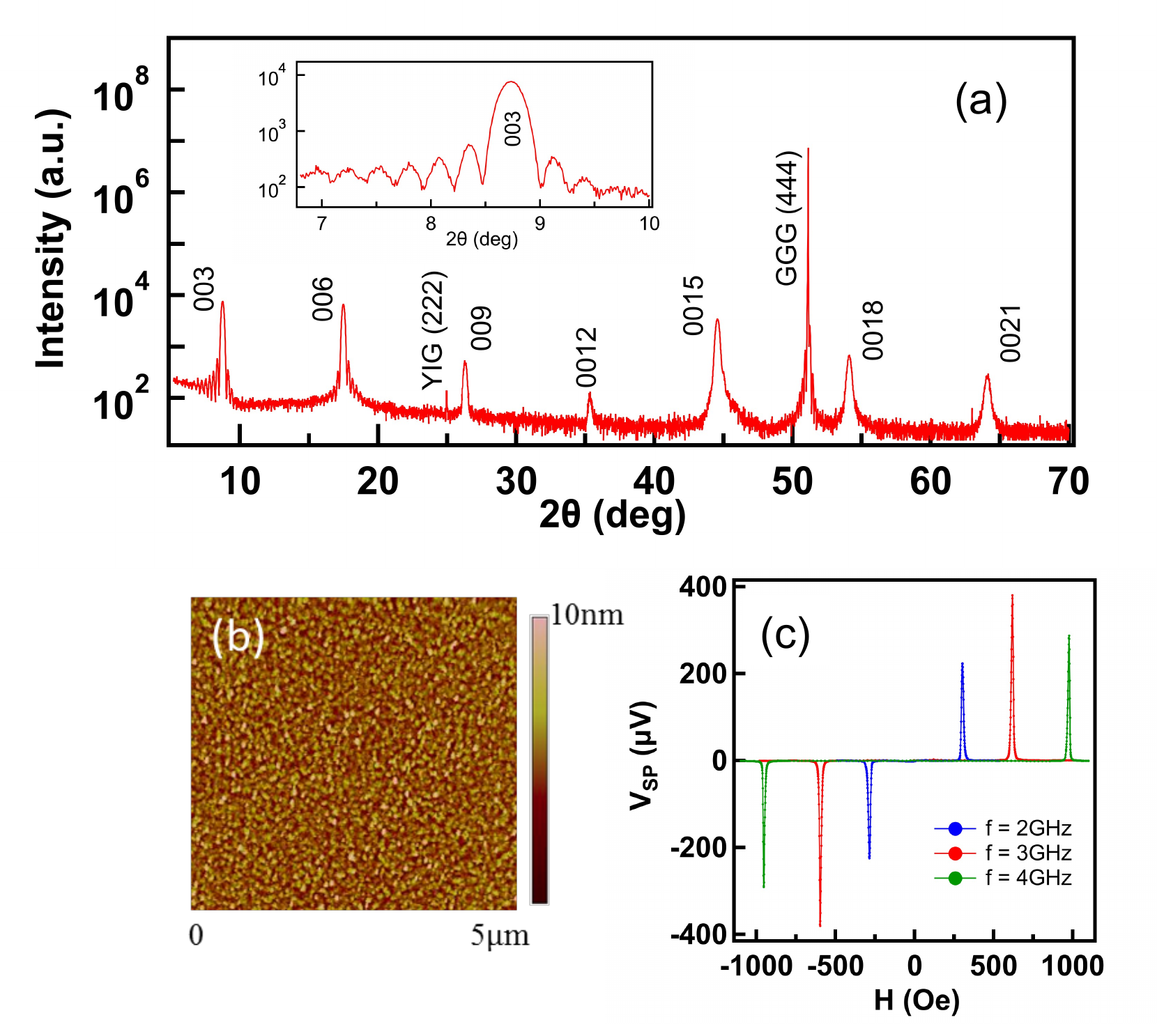}
\caption{\label{fig1} ((a) Semi-log $\theta-2\theta$ XRD scan of a YIG/\CBSTA~ (40 QL) sample which shows clear X-ray scattering peaks corresponding to the (003) to (0021) planes of \CBSTA. Inset: zoom-in view of the (003) peak curve shows pronounced Laue oscillations. (b) AFM image of a YIG/\CBSTA~(10 QL) bilayer over an area of $5 \mu \rm{m} \times 5 \mu \rm{m}$, which shows an rms roughness of 1.0 nm. (c) \Vsp~ vs. $H $ spectra of YIG/\CBSTA~ (10 QL) measured at microwave frequency 2 GHz, 3 GHz, and 4 GHz using 200 mW microwave power at room temperature.}
\end{figure}

We then carried out FMR-driven spin pumping measurements on YIG/TI samples (1 mm $\times$ 5 mm) by placing a microstrip transmission line on top of the sample and feeding it with microwaves. During the measurements, a DC field $H$ was applied in-plane along the microstrip line. At the YIG resonance condition, the interfacial dynamical exchange coupling \cite{Du:2013aa} between the precessing YIG magnetization and the charge carriers in the TI layer produces a pure spin current density \Js~ that flows at the interface into the TI layer. This spin current is converted into a two-dimensional (2D) charge current density \jc~ \cite{Rojas-Sanchez:2013aa,Rojas-Sanchez:2016aa} via spin-momentum locking in the surface states and/or a three-dimensional (3D) charge current density \Jc~  \cite{Tserkovnyak:2005aa,Heinreich2011,Mosendz} through the ISHE in the bulk states, resulting in spin pumping voltage signals (\Vsp) across the length of the sample. Figure 1(c) shows the observed \Vsp~ vs. $H$ spectra of a YIG/\CBSTA~ (10 QL) bilayer at three different microwave frequencies of 2 GHz, 3 GHz, and 4 GHz at room temperature. The observed spin pumping signals have all the expected hallmarks of a genuine spin pumping signal. For example, the signal changes sign when the polarity of the external magnetic field $H$ is reversed, as expected from either spin-momentum locking or the ISHE. The magnitude of the spin pumping signal also has a linear dependence on the microwave power.  Finally, we varied the geometry of the measurement to rule out possible artifacts\cite{PWang:2018} due to the Seebeck effect induced by lateral thermal gradients that might arise due to surface spin wave propagation in the YIG substrate (data shown in supplementary material). The robust spin pumping signal ($\sim 300 \mu$V) observed in our YIG/TI bilayer demonstrates the spin angular momentum transfer at the interface and provides an excellent platform to probe the underlying spin-charge conversion mechanism in the TI thin films.

To explore the comparative contributions from the surface and bulk states to the observed spin pumping signals, we tuned $E_F$ using two distinct methods: first, by electrical gating and, second, by varying the composition of the TI layers. In the latter case, $E_F$ varies for extrinsic reasons as the carrier density in the samples changes with the nature of the defects. We first discuss the variation of the spin pumping signal in an electrically gated sample. 

\begin{figure}
\includegraphics[width=3in]{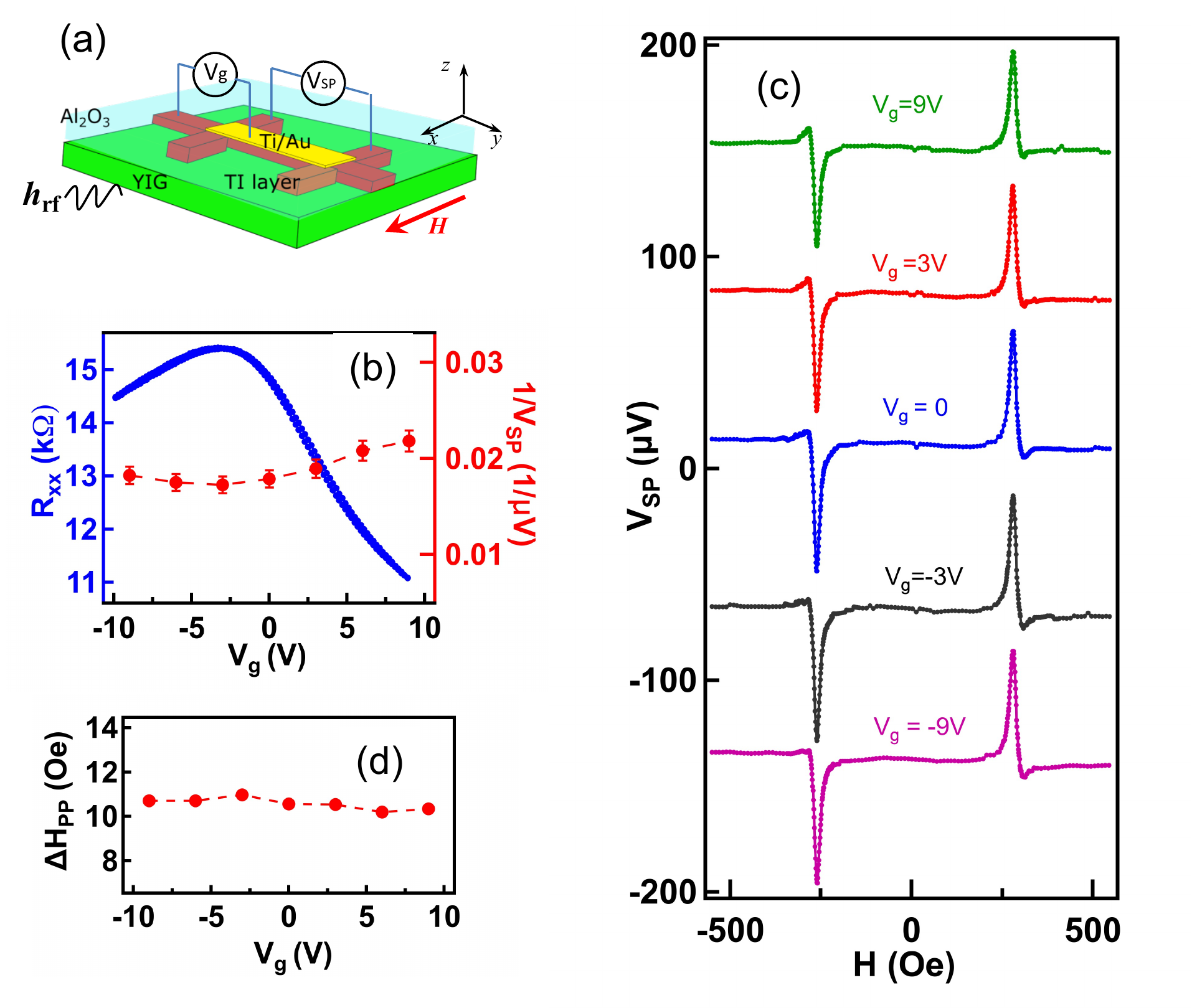}
\caption{\label{fig2} (a) Schematic of the device for electrical gating and spin pumping measurements. (b) Gate voltage dependence of the longitudinal resistance \Rxx~ (blue points) and inverse spin Hall signal $1/V_{sp}$ (red points); the latter is proportional to the spin Hall conductivity. (c)  Spin pumping signal $V_{\rm{SP}}$ vs. $H$ spectra at different gate voltages for the YIG/\CBSTA~(10 QL) sample measured at 50 K. The curves are offset for clarity. (d) Gate voltage dependence of the peak-to-peak FMR linewidth \DHpp}.
\end{figure}

Figure 2(a) shows the schematic of the devices for the electrical gating measurements. We patterned the YIG/\CBSTA~ (10 QL) bilayer into a standard Hall bar structure using photolithography and wet etching, followed by a 35 nm Al$_2$O$_3$ layer deposited by atomic layer deposition. A Ti(10 nm)/Au(80 nm) electrical contact vertically aligned with the Hall bar channel is defined on top of the Al$_2$O$_3$ layer to apply the gate voltage. To ensure the insulating nature and minimize the electrical leakage of the  Al$_2$O$_3$ layer, the measurement temperature is maintained at 50 K.

The left axis in Fig. 2(b) shows the gate voltage $V_g$ dependence of the longitudinal resistance (\Rxx) for a \CBSTA~ (10 QL) layer grown on a 20 nm YIG thin film measured at 50 K. At zero gate voltage, $E_F$ is expected to be close to the Dirac point in the bulk band gap (confirmed by the temperature dependence of the resistivity measurements shown below). As $V_g$ is swept to negative, $E_F$ moves down and passes the Dirac point which coincides with the maximum in longitudinal resistance (\Rxx $\sim 15.4 \rm{k}\Omega$ at $V_g = -3$ V) while the carrier type switches from n to p-type. The allowable gate voltage is limited by the electrical leakage in the Al$_2$O$_3$ layer and $E_F$ is expected to lie within the bulk band gap under the maximum applied gate voltage.

We now address the variation of \Vsp~ with $E_F$. Insulating systems with non-trivial topological character, such as topological insulators, have an approximate bulk-surface correspondence in the spin current similar to that found rigorously for the charge current in the quantum Hall state. The approximate nature of the correspondence is due to a finite spin relaxation rate, however this  rate can be slow compared with timescales relevant for spin transport. At a quantum Hall plateau, the calculated Hall conductivity is the same when calculated in the edge state picture and in the bulk picture, even though the relative contributions of edge and bulk states can vary in a non-universal fashion across a quantum Hall plateau \cite{Weis:2011aa}. For a surface-state with spin-momentum ``locking,'' the direction of the motion uniquely determines the spin orientations for the electrons or holes on the surface states. When a spin accumulation $\langle S \rangle$ with polarization $\vec{\sigma}$ is induced by FMR-driven spin pumping, a net momentum transfer is generated with the sign depending on whether the surface states are n-type or p-type. The opposite motion of electrons and holes produces the 2D charge current density \jc (A m$^{-1}$) with the same sign following $\vec{j_c} \propto \hat{z} \times \vec{\sigma}$ \cite{hasan2010,Kondou:2016aa}. From a bulk perspective, the collective motion of the full band produces a spin Hall conductivity $\sigma_S$ under the influence of a voltage, and thus $\sigma_S$ should be constant as $E_F$ changes within the gap \cite{Sahin:2015aa}. Calculations in either picture should produce the correct $\sigma_S$ when $E_F$ is within the gap. Figure 2 (c) shows the \Vsp~ vs. $H$ spectra of a YIG/\CBSTA~ (10 QL) bilayer measured at 50 K for different gate voltages and at 2.5 GHz microwave frequency. When $V_g$ varies from -9 to 9 volts, $E_F$ passes from below to above the Dirac point within the bulk band gap; note, however, that the sign of the observed spin pumping signal is unchanged and even the magnitude of the spin pumping signal is relatively constant.  Both these characteristics are broadly consistent with the bulk-surface correspondence picture.

We can extract the Fermi-level dependence of $\sigma_S$ from the spin pumping signal $V_{\rm{SP}}$ by noting that $\sigma_S \propto 1/V_{\rm{SP}}$. This is shown on the right axis of Fig. 2(b) via the $V_g$ dependence of $1/V_{\rm{SP}}$ (red points). This plot demonstrates the insensitivity of $\sigma_S$ to the Fermi level position within the bulk band gap. We also note that the spin pumping signal can be used to extract the Fermi-level dependence of the spin-charge conversion efficiency via the gate voltage dependence of $\frac{V_{\rm{SP}}}{R_{xx}}$.   This quantity also shows little variation with $E_F$ (see supplementary section). The magnitude of the generated 2D electrical charge current in the surface states is given by $j_c \approx \frac{2e}{\hbar} v_F \langle S \rangle$  [10, 13], where $v_F$ is the Fermi velocity. The 3D injected spin current density \Js (A m$^{-2}$) is proportional to $\langle S \rangle$ therefore, the spin-charge conversion efficiency $\frac{j_c}{J_s} \propto v_F$. Angle-resolved photoemission spectroscopy (ARPES) studies show that $v_F$ in the surface states of \BST~ thin films follows a linear dispersion with energy and keeps a constant value when $E_F$ is close to the Dirac point [30] which is consistent with our observations. We note that this trend of spin-charge conversion efficiency is in disagreement with recent Fermi level dependent spin Seebeck and spin-torque FMR reports in TIs \cite{Kondou:2016aa,Jiang:2016aa}. One possible reason for this discrepancy may come from the different mechanisms in tuning the Fermi level position in TI thin films. The electrical gating method used here for one YIG/TI bilayer avoids the potential issues coming from the variation of the electronic band structures \cite{Zhang:2009aa} and interface conditions between different samples and to a large extent provides a clear platform to probe the role played by surface states in spin-charge conversion. Another reason may be the difference in the fundamental measurement mechanisms between these probing techniques. The FMR-driven spin pumping process heavily depends on the short range exchange coupling at the YIG/TI interface. For the spin Seebeck effect, the long-range spin transfer mechanism \cite{Uchida:2011aa} may potentially complicate the picture and analysis. Figure 2 (d) shows the gate voltage dependence of the peak-to-peak FMR linewidth (\DHpp) for YIG/\CBST~ (10 QL) bilayers. \DHpp~ does not vary much with $V_g$, indicating that any possible electrical gating induced thermal heating is negligible.

\begin{figure}
\includegraphics[width=3in]{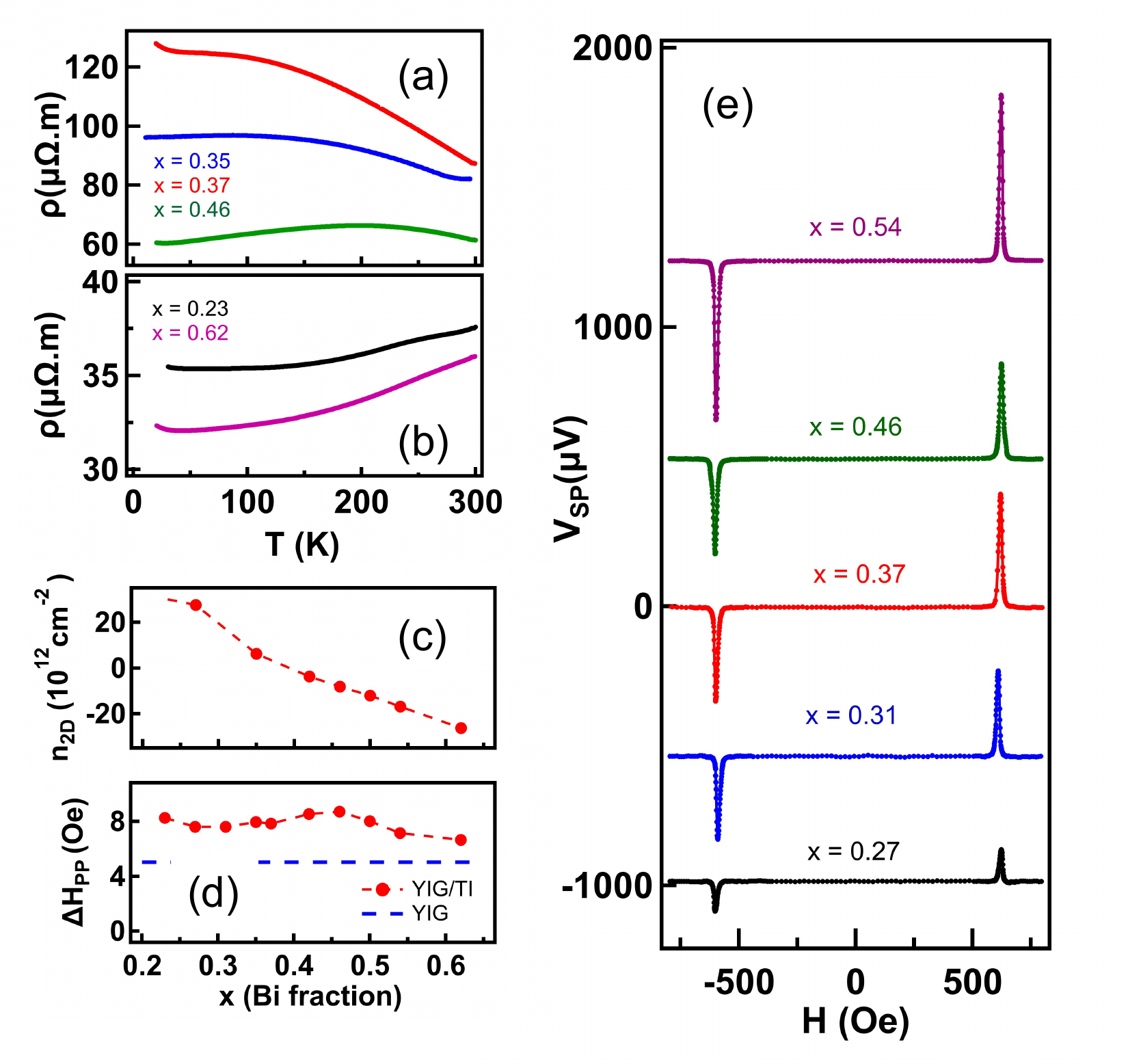}
\caption{\label{fig3} (a), (b) Resistivity of 10 QL \CBST~ thin films as a function of temperature. (c) Bi fraction dependence of 2D carrier concentration and (c) peak-to-peak FMR linewidth in YIG/\CBST~ (10 QL) samples. (Data measured at $T = 5$K. (e) \Vsp~ vs. $H$ spectra for YIG/CBST~ (10 QL) samples with different $x$ values measured at room temperature.}
\end{figure}

Since our present electrical gating methods do not allow us to vary $E_F$ all the way into the bulk valence or conduction bands, we also take another approach to tune $E_F$. Varying the composition ($x$) in \CBST~ (10 QL) thin films from 0.23 to 0.62 allows us to tune $E_F$ over a broader range, entering into both the conduction and valence bands. Figures 3(a) and (3b) show the temperature dependence of the longitudinal resistivity \Rxx~ of \CBST~ (10 QL) thin films at different $x$ values. For the \CBSTA~ sample, the resistivity dramatically increases by $\sim 50 \%$ from room temperature to 20 K, confirming the surface states dominate the longitudinal transport behavior [8, 11]. The Cr$_{0.08}$(Bi$_{0.35}$Sb$_{0.65}$)$_{1.92}$Te$_3$ and Cr$_{0.08}$(Bi$_{0.46}$Sb$_{0.54}$)$_{1.92}$Te$_3$ samples also demonstrate insulator-like behavior. The decrease in resistivity at low temperature when for $x = 0.23$ and $x = 0.62$ indicates a metallic behavior when Fermi level enters the bulk band \cite{Kondou:2016aa,Jiang:2015aa}. Figure 3 (c) plots the 2D carrier concentration $n_{\rm{2D}}$ obtained from the Hall measurements of different samples at $T = 5$K. To ensure a similar angular momentum transfer efficiency at different YIG/\CBST~ interfaces, Fig. 3(d) plots \DHpp~ vs. $x$ at 3 GHz at room temperature; this shows that \DHpp~ is almost constant $\sim 8$ Oe for $0.23 \leq x \leq 0.62$. The interfacial spin mixing conductance \cite{Tserkovnyak:2005aa} ($g_{\uparrow \downarrow \rm{YIG/TI}}$) is calculated from the line width broadening compared with bare YIG thin film (blue dash line); the values are in the range of $3 - 6 \times 10^{18} \rm{m}^{-2}$, comparable to those reported in YIG/transition-metal heterostructures \cite{Heinreich2011,Mosendz,Wang:2014aa}.

Figure 3 (e) shows the \Vsp~ vs. $H$ spectra of five YIG/\CBST~ (10 QL) samples measured at 3 GHz and 200 mW microwave power at room temperature. For the \CBST~ sample with $x = 0.37$, \Vsp $= 370 \mu$V. For $x = 0.31$ and $x = 0.46$, both the sign and the magnitude of the observed spin pumping signals are the same, confirming the insensitivity of the spin-charge conversion efficiency observed in electrical gating measurements when $E_F$ lies within the bulk band gap. As $x$ further increases or decreases, strikingly, we observe an enhancement of the spin pumping signal \Vsp~ to $\sim 600\mu$ V when $E_F$ enters the conduction band, and a dramatic decrease to $\sim 100 \mu$V when $E_F$ intersects the valence band. In a simple picture of a topological insulator, the spin-orbit correlations of states in the conduction band are opposite to those in the valence band, yielding a maximum spin Hall conductivity in the gap; however, in a more complex material with many bands near the gap there is no {\it a priori} requirement that the spin-orbit correlations of states near the conduction and valence band states edge should be opposite, as shown for Bi$_x$Sb$_{1-x}$ alloys [22].

We now discuss our experimental results in comparison with calculations of the spin Hall conductivity $\sigma_S$. Figure 4 (a) shows a qualitative determination of $E_F$ relative to the Dirac point inferred from the 2D carrier concentrations determined using the Hall effect. To make a direct comparison with theoretical calculations, we plot the Fermi-level dependence of the spin Hall conductivity $\sigma_S \propto 1/V_{\rm{SP}}$.  Figure 4(b) shows the variation of $1/V_{\rm{SP}}$ with the Bi fraction. This Fermi energy dependent spin Hall conductivity can be qualitatively explained by the ``full-band'' contributed spin Hall conductivity in \BST~ along with calculations of the spin-orbit correlations in the conduction and valence bands\cite{Zhang:2009aa}. We calculate $\sigma_S = \sigma_{yx}^z$ for \BST~ using a technique previously described \cite{Sahin:2015aa}. The electronic structure is parameterized by a tight-binding model of the material, and the Berry curvature is directly evaluated. We then sum up the Berry curvature for all occupied states up to the Fermi energy, which yields a non-vanishing value within the TI energy gap. The results of the calculation are shown in Fig. 4(c). As the Bi fraction is increased from $x = 0.31$, $E_F$ shifts up from the bulk valence band and enters the bulk energy gap. Both theory and experiment show a marked decrease in the spin Hall conductivity in this regime. When the Bi fraction is increased further and $E_F$ crosses the bulk gap while coincident with the Dirac surface states, the spin Hall conductivity from this calculation does not change. This is because there are no bulk states within the TI energy gap. Again, both experiment and theory show qualitatively consistent results in this regime. We also note that these observations are consistent with our observed electrical gating measurements shown in Fig. 2(b). As the Bi fraction is increased even further ($x \geq 0.5$), $E_F$ enters the bulk conduction band. Here, we find some disagreement between theory and experiment for the two samples measured in this regime. However, the carrier concentration has been measured at a different temperature than the spin current. Theoretical calculations that assume a hole concentration 2.5 times smaller at room temperature than the low-temperature measured value, and an electron concentration 2.5 times larger at room temperature than the low-temperature measured value, are in good agreement with the experimental spin signal (Fig. 4(d)). There are several possible sources for this difference, including, {\it e.g.}, the low-temperature freeze out of electron traps. 

\begin{figure}
\includegraphics[width=3in]{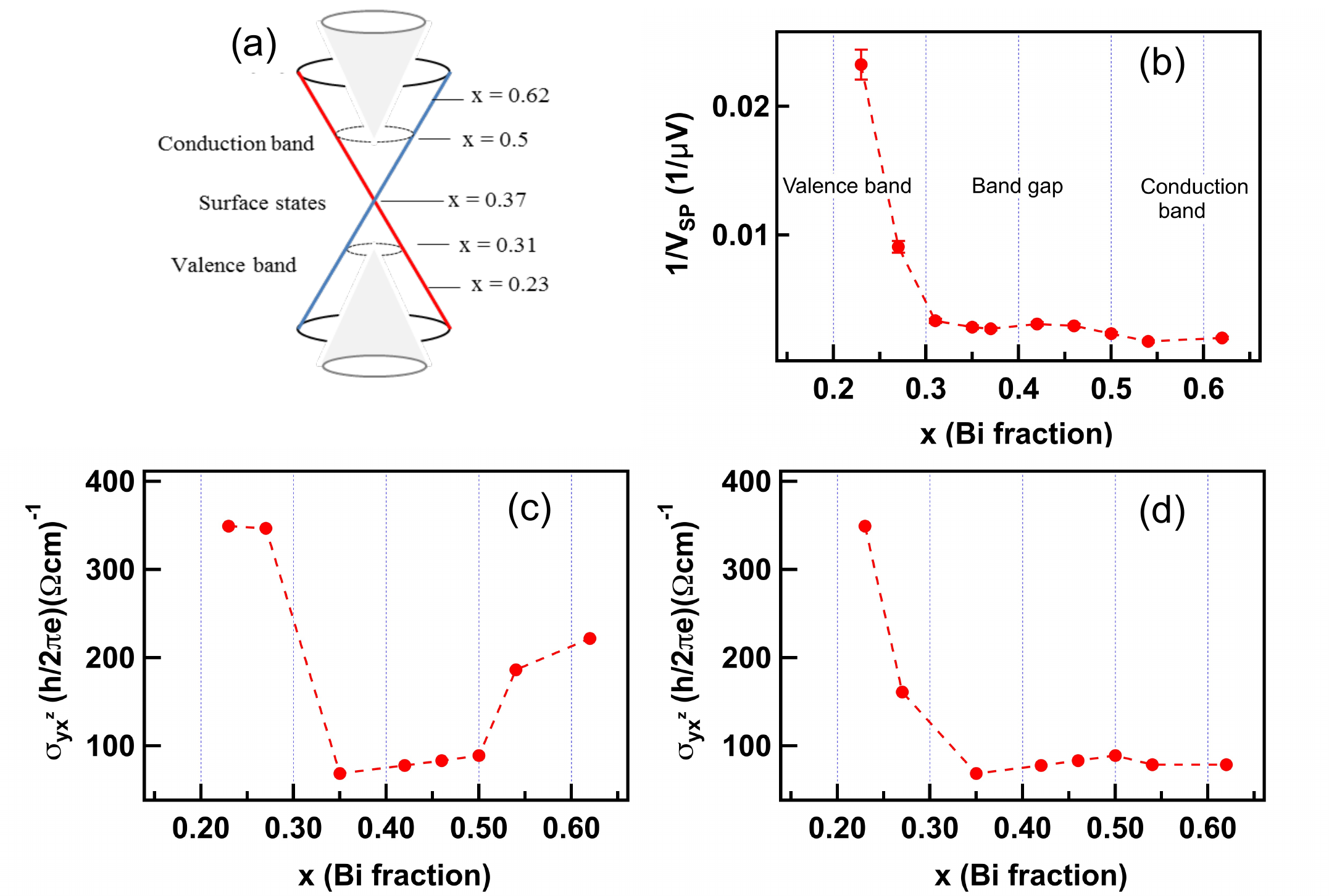}
\caption{\label{fig4} (a) Qualitative schematic electronic band structures of \CBST~ with the Fermi level position at different $x$ values. (b) Bi fraction dependence of the spin Hall conductivity characterized by $1/V_{sp}$. (c) Calculated spin Hall conductivity for \BST~as a function of Bi fraction using carrier density measured from Hall effect. (d) Calculated spin Hall conductivity for \BST~as a function of Bi fraction using a hole carrier density 2.5 times smaller when the chemical potential is in the valence band and an electron carrier density 2.5 times larger when the chemical potential is in the conduction band.}
\end{figure}

In summary, we report the variation of the spin-charge conversion efficiency with the Fermi level position tuned by both electrical gating and varying the composition in TI thin films. The opposite trends of the observed spin pumping signals when the Fermi level enters the conduction and versus valence bands demonstrate the similar spin-orbit correlations in the conduction and valence bands due to the complex electronic structure of the material. Despite this, the bulk-surface-state correspondence for topological insulators is preserved within the gap, in which the spin Hall conductivity is insensitive to the Fermi level. This result points to the important ability to tune characteristics of topological-insulator based spin functional devices by making use of both surface state and bulk bands.

\begin{acknowledgments}
HW, JK, MF, NS, and MW  acknowledge support from C-SPIN, a funded center of STARnet, a Semiconductor Research Corporation (SRC) program sponsored by MARCO and DARPA. WY and NS are supported by SMART, one of seven centers of nCORE, an SRC program, sponsored by National Institute of Standards and Technology (NIST). AR and NS acknowledge partial support from ONR (N00014-15-1-2370).  C\c{S} and MEF acknowledge support from the Center for Emergent Materials, an NSF MRSEC under Award No. DMR-1420451. TL and MW acknowledge support from the U.S. Department of Energy (DE-SC0018994) and the National Science Foundation (EFMA-1641989).
\end{acknowledgments}

%


\end{document}